\documentclass[letterpaper,twocolumn, 10pt]{article}
\usepackage{usenix-2020-09}
\usepackage{graphicx}

\usepackage{url}
\usepackage{xcolor}
\usepackage{tabularx}

\begin{document}

\date{}

\title{\Large \bf  Voice App Developer Experiences with Alexa and Google Assistant: \\ Juggling Risks, Liability, and Security}

\author{
{\rm William Seymour}\\
King's College London
\and
{\rm Noura Abdi}\\
Liverpool Hope University
\and
{\rm Kopo M. Ramokapane}\\
University of Bristol
\and
{\rm Jide Edu}\\
University of Strathclyde
\and
{\rm Guillermo Suarez-Tangil}\\
IMDEA Networks Institute
\and
{\rm Jose Such}\\
King's College London \&\\
Universitat Politecnica de Valencia
}

\maketitle

\begin{abstract}
Voice applications (voice apps) are a key element in Voice Assistant ecosystems such as Amazon Alexa and Google Assistant, as they provide assistants with a wide range of capabilities that users can invoke with a voice command. Most voice apps, however, are developed by third parties---i.e., not by Amazon/Google---and they are included in the ecosystem through marketplaces akin to smartphone app stores but with crucial differences, e.g., the voice app code is not hosted by the marketplace and is not run on the local device. Previous research has studied the security and privacy issues of voice apps in the wild, finding evidence of bad practices by voice app developers. However, developers' perspectives are yet to be explored. 

In this paper, we report a qualitative study of the experiences of voice app developers and the challenges they face. Our findings suggest that: 1) developers face several risks due to liability pushed on to them by the more powerful voice assistant platforms, which are linked to negative privacy and security outcomes on voice assistant platforms; and 2) there are key issues around monetization, privacy, design, and testing rooted in problems with the voice app certification process. We discuss the implications of our results for voice app developers, platforms, regulators, and research on voice app development and certification.
\end{abstract}

\maketitle

\section{Introduction}
\label{sec:introduction}

In the decade since the arrival of Amazon Alexa and Google Assistant, voice assistants have significantly increased in functionality and accuracy and are now widely integrated into smartphones, TVs, speakers, and many other devices. The nature of the technology---always-on and always-listening---has sparked significant interest in the privacy of voice assistants and people's associated perceptions, with a growing body of work investigating these concerns both qualitatively ~\cite{lau2018alexa,10.1145/3359161,VIMALKUMAR2021106763,abdi2019more,10.1145/3491102.3517510} and quantitatively ~\cite{abdi2021privacy,kennedy2019can,dong2020your,cho2019hey}.

At the same time, mirroring societal and legislative attention around the moderation of online platforms such as the EU Digital Services Act~\cite{DSA} and the UK Online Safety Bill \cite{Online_Safety_Bill}, researchers have started to investigate how third-party voice apps are approved for distribution on voice stores by voice assistant platforms (\emph{skills} for Amazon Alexa and \emph{actions} for Google Assistant). To date, studies have begun to map out the opaque and poorly documented use of human and automated testing of voice apps by voice assistant platforms~\cite{10.1145/3478101}, criticizing the effectiveness of the voice app certification process~\cite{cheng2020dangerous,10.1145/3543829.3544513}. Focusing on the largest platform in the space (Amazon Alexa), this has led to a consensus amongst researchers that current efforts are insufficient, with studies continuing to show that violation of content policies~\cite{cheng2020dangerous, young2022skill}, privacy regulation violations~\cite{edu2021skillvet}, poor separation of adult-and child-focused voice apps~\cite{le2022skillbot}, and circumvention of the permission system~\cite{guo2020skillexplorer} are prevalent on the Alexa voice app store despite responsible disclosure efforts by researchers to Amazon~\cite{edu2022measuring}.

Thus far, this line of research has focused on \textit{user} perspectives and Amazon and Google as platform owners; to the best of our knowledge, none have considered the experiences of third-party \textit{developers} of voice apps for Alexa and Google Assistant. Rectifying this is crucial, given that developers are the group with primary control over voice apps, content, and privacy policies that appear on the platform. Understanding the circumstances that lead to the creation of problematic voice apps is, therefore, vital in understanding what can be done to improve the health of the wider ecosystem.

To investigate this and explore the reality of developing for the platform, we conducted semi-structured interviews with 30 voice app developers with varying levels of expertise about the challenges they face in creating skills and actions. Starting with the development process, we then asked more focused questions on data \& privacy, security \& testing, certification, their relationship with the major platforms, and the developer economy. Through thematic analysis of the transcripts, we answer the following research questions:

\begin{enumerate}
    \item[RQ1] What are the main challenges faced by voice app developers that are specific to voice app development?
    
    \vspace{-8pt}
    
    \item[RQ2] How do these challenges relate to security and privacy on voice assistant platforms?
    
    \vspace{-8pt}
        
    \item[RQ3] How do these challenges and developer responses to them differ across developers with varying levels of experience?
    
        \vspace{-5pt}
\end{enumerate}

\noindent And, in so doing, we make the following main contributions:
    \vspace{-5pt}

\begin{itemize}
    \item We identify key issues around monetization, privacy, design, and testing and show how many of these are rooted in problems with the voice app certification process.
        \vspace{-5pt}

    \item We show how issues around, e.g., liability and certification are linked to negative privacy and security outcomes on voice assistant platforms.

        \vspace{-5pt}

    \item We discuss the implications of our results and make recommendations for developers, regulators, and researchers on voice app development and certification.
\end{itemize}

Key findings include developers' feeling that the risk and liability of developing skills had been pushed onto them with little support, widespread reuse of privacy policies, and the extreme difficulty of monetizing voice apps. The poor discoverability of voice apps was felt by all developers and intensified existing problems around design, user experience, and monetization. Certification, the key point where privacy and security issues can be identified before voice apps become publicly available, was seen as arbitrary, inconsistent, and open to influence from internal contacts, leading several developers to create mitigation strategies to reduce the coverage of certification checks. We then discuss the implications of these findings for privacy and security on Alexa and Google Assistant platforms, regulatory compliance, the developer economy, and for future research into the security and privacy of voice assistant platforms.

\section{Background}
\subsection{Voice App Architecture}

\begin{figure}[h]
    \centering
    \includegraphics[width=0.8\columnwidth, trim=1 1 1 1, clip]{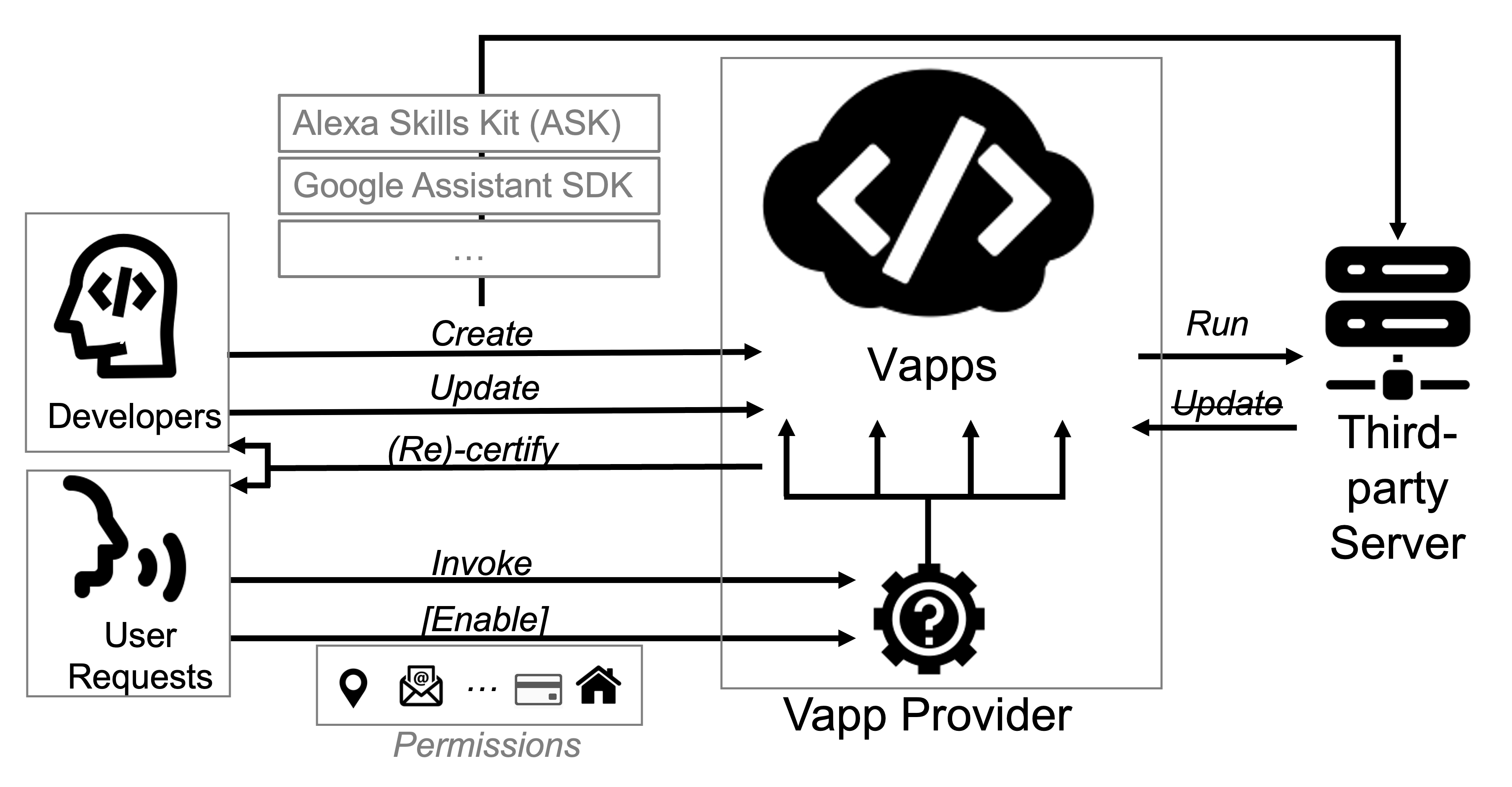}
    \caption{Overview of the interaction between developers and voice assistant architecture components.}
    \label{fig:voice apps}
\end{figure}

Voice assistant platforms allow developers to create voice apps using dedicated APIs and tools for which they offer comprehensive documentation.
For example, Amazon maintains the Alexa Skills Kit (ASK) and Google provides the Assistant SDK. These complex ecosystem components allow developers to execute code, namely, skills or actions, in response to user requests. Figure \ref{fig:voice apps} illustrates the position of voice apps within the voice assistant ecosystem.

Unlike smartphone apps, voice apps do not run any code on user-controlled devices. Instead, audio from the assistant, such as the audio captured by an Amazon Echo smart home device, is streamed to the associated platform (e.g., Amazon Alexa). The platform then transcribes the audio and attempts to match it to the best voice app to serve the user command via a pipeline of machine learning models running in the cloud to process voice commands. Moreover, the platform will also determine which of the voice app's `intents' (similar to functions) should be invoked. The code for the voice app itself is run in the cloud either as an Amazon Web Services (AWS) Lambda function, on Google Cloud, or on a server controlled by the developer. An API is provided by the platform to ensure that the voice app's endpoints can be correctly reached, and responses are piped back through the platform and transformed into speech, which is then sent back to the device (e.g., Amazon Echo) to be played for the user. This cloud-based architecture allows developers to easily update voice apps and scale their capabilities to accommodate changing demands. 

\subsection{Accessing user information} 
Both Amazon and Google provide developers with ways to programmatically request access to user information, such as location data, name, and contacts via API calls. In most cases, these data are provided by the platform rather than the user's device. To obtain consent from users, platforms display the types of data that the voice app will request access to in their marketplace, as is the case with smartphone apps. Users must explicitly grant access to these data when enabling these voice apps. 

While the use of the permission framework is normally the only approved means of accessing user data, prior work has found that some developers request information directly from users through the conversational interface. In this case, platforms are unable to perform runtime checks, and do not have a mechanism to ensure that developers have declared the intention to collect specific data~\cite{edu2021skillvet}. 

\subsection{Voice App Certification}
Third-party voice apps on both Alexa and Google Assistant have to undergo and pass a certification process to ensure compliance with platform policies and security and privacy requirements~\cite{amazon-policy-testing, google-policy-testing}. 
If and when voice apps are approved, they become available to users through the store, and users can ask the assistant for them by name. Once certified, developers cannot make changes to the interaction model, including the utterances that a user can say to the app, without first submitting the app for re-certification. However, updates to the back-end of a voice app---i.e. the part that generates responses to user input---{\em do not require re-certification}~\cite{cheng2020dangerous}.

\subsection{Voice App Security and Privacy Issues }
\label{subsec:issues}
A key area of research on voice apps and assistants centers around privacy and security issues, including several unique to voice apps. We offer a summary of those issues as a way of introduction and background, but details can be found in Section~\ref{sec:related}. For instance, due to the way discoverability works in the ecosystem, voice app \emph{impersonation} is possible, where the invocation name of a voice app can be selected to match that of the target voice app or be similar enough phonetically~\cite{kumar2018skill}. Other 
 security issues include voice squatting, a voice app faking or ignoring the  termination command and continuing to operate stealthily~\cite{SRLabs_Squatting, zhang2019dangerous}; and voice masquerading~\cite{zhang2019dangerous}, a voice app pretending to hand over control to another voice app to eavesdrop user conversations.

When it comes to privacy, large-scale static~\cite{lentzsch2021hey, edu2021skillvet} and interrogative~\cite{guo2020skillexplorer, young2022skill} traceability analyses of voice apps reveal bad privacy practices and misleading privacy policies. Research has also revealed a prevalence of voice apps gathering and storing medical data in clear violation of platform policies~\cite{shezan2020verhealth}. Whilst current voice app ecosystems only support permission-based access control for sensitive data, this has been shown to be insufficient to control how voice apps use data once they have access~\cite{guo2020skillexplorer,10.1145/3412383,edu2022measuring} Finally, account linking can be exploited to obtain sensitive user information~\cite{edu2021skillvet, edu2022measuring}.

These security and privacy issues are often facilitated as the code of a voice app can be modified at will by third-party developers, bypassing the skill certification process~\cite{cheng2020dangerous, lentzsch2021hey}, as it may reside on remote web services under their control~\cite{su2020you, SRLabs_Squatting}. The certification process for voice apps has been found to be flawed and ineffective, leaving space for many malicious or policy-violating apps to be published~\cite{cheng2020dangerous,lentzsch2021hey}.

The issues mentioned above are compounded by users' limited comprehension of the voice app ecosystem and the intricate dynamics of the information flow within it~\cite{abdi2019more,abdi2021privacy, lau2018alexa,10.1145/3170427.3188448, VIMALKUMAR2021106763,10.1145/3491102.3517510} While previous research has extensively investigated the security and privacy concerns associated with voice apps, the motivations and circumstances behind the development of problematic voice apps remain unexplored.

\section{Methods}\label{sec:methods}
To answer our research questions, we designed a qualitative user study based on semi-structured interviews with voice app developers that we analyzed using thematic analysis. All parts of the study were approved by our institution's IRB and supplementary material for the study is available at \url{https://osf.io/yj7es} to aid replication and promote open science.

\subsection{Recruitment}
\label{subsec:recruitment}
We recruited 30 developers of Amazon skills and Google actions via publicly available contact information on the respective stores, as well as in-person through a local voice app developer network in London. Interviews were conducted in person or online via Microsoft Teams, depending on the location of the participant. Following best practices for sampling in qualitative research for security and privacy studies~\cite{redmiles2017summary}, we prioritized achieving a balance of demographics relevant to our focus, such as: platform developed for, types of skills developed, number of skills developed, and diversity of experience. 

We classified developers as either hobbyists or professionals based on whether they had ever developed a voice app as part of their main employment --- yes for professionals, no for hobbyists. This is because when choosing how to categorize developers in terms of their experience, we felt it important to decouple this definition from characteristics such as the number of voice apps developed or whether they had been monetized, as these were distributed across the participant pool owing to factors laid out in Section~\ref{sec:results}---e.g., monetization of voice apps ranged from a few dollars a month through to tens of thousands.

As it can be seen in the full summary of participants given in Table~\ref{tab:demographics}, our participants covered a broad range of voice app categories and levels of experience, with half having developed skills or actions in a professional capacity. Some participants, particularly the hobbyists, had developed a single voice app, with a clear group of professional developers who had developed more than 6 voice apps. The voice apps developed by our participants covered a wide range of categories (11+), with games being the most common category. 

\begin{table}
    \centering
    \footnotesize
    \begin{tabular}{l|r|l|r}
    \hline
    \multicolumn{2}{l|}{Number of Developers:} & \multicolumn{2}{l}{Number of voice apps developed:} \\
    \hline
    Professional & 15 & 1 & 15 \\
    Hobbyist  & 15 & 2-4 & 6 \\
    &&4-6 & 2 \\
    &&6+ & 7\\
   \hline
   \multicolumn{4}{l}{Number of developers and voice app categories:} \\
    \hline
    Gaming & 8 & Productivity & 2 \\
	Health and Well-being & 3 & Lifestyle & 3 \\
	Education & 3 & Travel \& Transportation & 1 \\
	Weather & 1 & Music \& Audio & 3\\
	News & 1 & Novelty \& Humor & 2 \\
	Food \& Drink & 3 && \\
    \hline
    \end{tabular}
    \caption{Participant Demographics.}
    \label{tab:demographics}
\end{table} 

\subsection{Interviews and Protocol}
The interview protocol was semi-structured, which is known to be very good at allowing interviewees to fully express their opinions and encouraging them to provide more useful information, while giving the interviewers reliable, comparable qualitative data~\cite{knott2022interviews}. The full version of the initial script used can be consulted in the ``Interview Outline'' file in the repository linked at the beginning of this section. It contains questions on the following topics:

\vspace{-5pt}

\begin{itemize}
    \item Motivation for developing voice apps, and general level of experience.

    \vspace{-5pt}

    \item Decisions around voice assistant platforms, hosting arrangements, and experiences with those platforms.

    \vspace{-5pt}

    \item Privacy and security arrangements for voice apps, including data collection, requested permissions, and privacy policies.

    \vspace{-5pt}

    \item Experiences publishing voice apps (certification, updates, etc.)
\end{itemize}

As standard in semi-structured interviews, beyond the questions in the script, researchers asked follow-up opportunistic questions in each interview, as appropriate, to fully draw out participants' experiences. For instance, this included interviewers restating and summarizing the interviewees' answers to confirm their opinions as well as formulating new questions based on interviewees' answers~\cite{westby2003asking}. Interviewers also used open-ended questions, avoiding dichotomous questions that only led to two opposing answers, leading questions, or why questions~\cite{westby2003asking}.

In order to refine the interview protocol, we used the first three interviews as pilots, taking time to refine the interview questions based on participants' responses and realign them with the research questions. For example, we changed the order of some of the questions, such as current questions 3 and 4, by first asking about the platforms used and then the development, as this would lead to a more natural conversation flow. The three pilot interviews were excluded from the final data set for analysis, leaving 27 transcripts for full analysis (referred to as P1--P27 in-text). Participants received a \$20 gift card upon the completion of the interview.

\subsection{Analysis} \label{sec:analysis}
Interviews were recorded and transcribed by researchers before being analyzed using thematic analysis~\cite{braun2006using}. On average, each interview took 30 minutes, with a total of circa 900 minutes worth of recorded audio. After familiarizing themselves with the data, two researchers coded the first transcript before discussing and arriving at the initial codebook. They then refined these codes in the second interview, discussing and agreeing on the final codebook that was applied to the remainder of the interview transcripts.

Following best practice, interviews were then conducted in batches, with researchers considering the extent of ``rich, complex and multi-faceted'' results being obtained, before making an in-situ decision based on the data obtained to keep increasing the sample size~\cite{sandelowski2001real}. However, there is widespread acknowledgment that raw sample size is not the main factor affecting data quality in qualitative research~\cite{braun2021saturate,sandelowski2001real}, so we also developed the composition~\cite{vasileiou2018characterising} of the sample to be as diverse as possible with regard to the aspects crucial in this domain as detailed in Section~\ref{subsec:recruitment}, e.g.: by continuing to increase the sample size until we had as many professional developers, who were much more difficult to recruit, as hobbyists.

After all of the interviews were completed, two researchers reviewed the coded transcripts and searched for and reviewed potential themes. These were incrementally written up as brief notes, an initial summary, and then as a report. Throughout the analysis process the entire research team met regularly to review progress and resolve disagreements. 

When reviewing the transcripts, a correlation was noted between the depth of descriptions given by participants and their level of development experience; in many cases, hobbyist developers used ambiguous terminology and/or sparse answers that became clearer when reviewed alongside the richer context and more precise definitions given by professionals. As such the reporting of quotes is weighted towards the latter group for clarity even though the concepts they describe were encountered more widely. Where the results are specific to a particular group of developers or only one of the voice assistant platform studies, we indicate this in the text ---e.g. by referring to \textit{skills} and \textit{actions} rather than \textit{voice apps}. Reported quotes have been edited to remove filler words and guggles (e.g., `umm', `like', `ah-ha'), with [..] used to indicate where quotes have been condensed for brevity.

\section{Findings}\label{sec:results}
\subsection{Money, Power, and Platform Influence}
\subsubsection{Challenges Around Monetization}\label{results:monetisation}
Almost all the hobbyist developers we interviewed had not made money from developing voice apps, although many were open to the possibility: \textit{``I was planning to monetize, but [the skills] I built it didn't generate any income''} [P08]. Instead, these developers were much more likely to be `paid' with non-monetary items such as AWS credits --- which often equated to them not having to pay for the privilege of developing skills --- or with other items: \textit{``I did it for some for some bounties to get some T-shirts or socks from Amazon''} [P16].

Those who had managed to generate regular incomes described the significant challenges involved. On other online platforms, developers commonly earn a living through advertising and payments for or within-apps. As advertising is prohibited on Google Actions and was only recently allowed in some cases for Alexa skills after the interviews had concluded,\footnote{\url{https://developer.amazon.com/en-GB/docs/alexa/custom-skills/policy-requirements-for-an-alexa-skill.html} (Section 5, accessed 16/03/2023).} this left in-app payments as the obvious choice.

But professional participants complained that the functionality for in-app purchasing had been neglected on both platforms, and was \textit{``pretty painful from a usability point of view''} [P21], having not received the refinements necessary to prevent it from being a frustrating and/or scary experience for customers. This, in turn, meant that the `freemium' model that supports many apps on other platforms did not work for voice apps: \textit{``you lose about two-thirds of your users at the skill connection, then lose another two-thirds in your payment flow. And these are people who have already said yes to the fact that they want to purchase''} [P21].

The above, coupled with generally lower user counts meant that the economies of scale leveraged by e.g. mobile apps were considered infeasible. On this note, one participant suggested that concerns around the \emph{security} of being able to spend money via voice were also partly to blame; that other people being able to use your voice assistant felt \textit{``like someone just getting hold of your phone''} [P22]. This helps explain why users may be more hesitant to make purchases through voice vs smartphone apps, and suggests a direct connection between issues raised by developers around fundamental difficulties with monetization on voice assistant platforms and fears around voice assistant security that have been captured in the literature~\cite{abdi2019more}.

\subsubsection{The Intersection of Money and Power}\label{results:moneyandpower}
Another key option for making money on the Alexa platform was the Developer Rewards Program, where Amazon sends payments to developers of the top-performing skills on the platform. Whilst originally representing a considerable portion of overall developer compensation, it has since been reduced in size to the point where it no longer supports full-time development:

\vspace{-8pt}

\begin{quote}
    \textit{``I think that at its peak, we used to get nearly 30 grand a month out of developer rewards, so it was quite a sizable amount of money. And now that's super tailed off. Despite having a lot more usage [now], I think you're lucky to get over a thousand for it''} [P21]
\end{quote}

The final means of earning money was responding to bounties offered for developers of voice apps in sectors of interest or implementing beta functionality in exchange for the promotion of their skills, although the rewards were seen as being too low to be sustainable as an income (we discuss this more in Section~\ref{results:privacy}). In practice, the de jure or de facto lack of monetization had driven away many smaller developers who might have gone on to develop the next big thing for Alexa:

\vspace{-8pt}

\begin{quote}
    \textit{``It's a car crash, a total car crash. And it's part of the reason why I moved away from running my own business and took the job at the bank. [..] [At a voice gaming conference] there was a guy there talking about his Alexa games [..] and after 4 or 5 years he's making dozens of dollars a month''} [P22]
\end{quote}

The total control of these means of monetization by platforms highlights the power they wield over developers. Especially for those dependent on the development of voice apps for their jobs, \textit{``the platforms are very powerful players in the market, we're entirely reliant on them for access to our customers. Having multiple platforms helps mitigate the imbalance in the relationship''} [P21]. In relation to this, several participants mentioned the decision by Google to deprecate conversational actions for Google Assistant in order to focus on app actions (where voice commands perform actions within apps on the Google Play Store).\footnote{\url{https://developers.google.com/assistant/ca-sunset} (accessed 16/03/2023).}

To mitigate the power held by platforms, some utilized middleware applications that allow the same business logic to be deployed across multiple voice assistants. This software also enabled voice apps to mix and match functionality between providers by, for instance, swapping out the product used for intent matching, and to be deployed across a range of other chat-based platforms including WhatsApp and Facebook Messenger. For small-scale developers, the neglect of in-skill and platform-provided monetization options increases the incentives for  developers to pursue `riskier' monetization strategies, enabled by platform negligence in identifying and removing skills with non-compliant or illegal privacy practices as described in Section~\ref{subsec:issues} and further explored with our participants in Section~\ref{results:privacy:use-policies}.

\subsubsection{Indirect Monetization} In contrast, the most effective means of monetizing voice apps was indirect. For larger companies looking to divert support requests that would usually be handled by, for instance, a call center through a voice app, the effective returns were much higher: \textit{``if we can channel half a percent of users who would have called to Alexa, we're saving thousands a day''} [P23]. In practice, the exact manifestation of this kind of cost-saving voice app depends on the particular situation being considered:

\vspace{-8pt}

\begin{quote}
\textit{``Sometimes a chat bot on a website is much better than [a voice app] because the user is already on the website [..] and they're looking up the contact number for someone so they can call. No, we don't want them on the call center. That costs company [X] pounds per user per call [..] whereas other things would be, actually, what we want is we want a device-free experience. So for example, it's not necessarily now about working out what's wrong with my device. It's working out how do I fix it, what do I need to do [..] and at that point, we'll go, well, actually, you don't want to be having your phone up and scrolling through, you just want Alexa to tell you''} [P23].
\end{quote}

And in some cases, it was suggested that the most profitable option was not to make a voice app at all, instead opting for what was described as `voice SEO': \textit{``we're [often] saying the smartest thing that you could possibly do with your money right now is investing in making sure that Google [Assistant] and Alexa are saying the right answers to your [product] rather than whatever else''} [P23].

\subsection{Privacy, Risk, and Liability}\label{results:privacy}
\subsubsection{Pushing Risk and Liability onto Developers} The need for competition was also highlighted by developers feeling unfairly burdened with liability and risk as a defining feature of their relationships with platforms. A prime example was how Amazon offered small bounties for skills in sectors of interest or offered early access to new functionality that developers could implement in exchange for the promotion of their skills. For less experienced developers this could appear as a gesture of goodwill: \textit{``they also even give us advance notice of certain things that they're going to develop in the future before even the wider public knows about it.''} [P20], but those more experienced were less positive: 

\vspace{-8pt}

\begin{quote}
    \textit{``There's a quid pro quo thing and they were like, do this and we'll promote you [..] so I asked them, well, what research have you done around this? Have you seen users asking for this? And they said, oh at Amazon, we see it as our job to invent on behalf of the customer. So we kind of come up with ideas and then we'll get partners like you to implement it. And then you will then see whether it works or not. [..] I heard from people that did go into that beta that it was a complete nightmare. The product didn't work at all. And it basically got put on ice and never formally launched.''} [P21]
\end{quote}

Those who did opt to implement beta functionality sometimes encountered additional problems when this had not been communicated to those responsible for certification: \textit{``depending on where it is in the beta functionality will depend on how polished it is. So sometimes we get really buggy beta functionality for Amazon. That then goes into certification process. And the certification team haven't been briefed on the functionality. So you then deploy, you then send it to certification team and they go, what is this? [..] we've had absolute nightmares with that.''} [P23].

Further reinforcing the importance of platform choice, some developers described the way that Amazon was seen to be predatory around third-party developers `competing' with it, leading to reticence on the part of developers to create skills for the platform: \textit{``they can be quite a predatory company, they can come into sectors and then through a combination of pricing, reduced cost of delivery through technology, they can just dominate a sector''} [P22].

This phenomenon was partially enforced through the certification process, and included an instance where a developer had a skill certification held up and ultimately rejected as a new item was added to the certification policy. Shortly afterward, an Amazon company launched a skill with the same feature:  \textit{``they changed the policy, they added a policy point that it wasn't allowed. And then about six weeks later [an Amazon-owned company] launched a [feature]-based skill''} [P21].

Those more content with their relationship with Amazon tended to either be developing skills for fun/to satisfy their curiosity or in agencies/companies established enough that either clients or Amazon were paying \textit{them} to develop for their platforms, and thus not dependent on direct income from their voice apps.

\subsubsection{Personal Data Use and Privacy Policies} 
\label{results:privacy:use-policies}
For privacy (and specifically privacy policies), while all participants understood that a voice app's privacy policy was their responsibility, for hobbyists it was often conveyed as a requirement imposed by platforms rather than required by regulations such as the GDPR~\cite{GDPR16}. At all levels, participants avoided writing their own privacy policies, with a variety of strategies depending on the situation: \textit{``I found some really good privacy policies online and copied them.''} [P06], \textit{``My approach there was always just to send a link to [the client's] privacy policy, just the one that's linked on their website.''} [P22]. Several participants reported using a Google-provided template when making Google actions, which could then be re-used in future voice apps:

\vspace{-8pt}

\begin{quote}
\textit{``I think that the privacy policy was originally a template that we got from somewhere. In fact, I think the original template even came from Google [..] and so I think we took that and then I think we adapted it over the years.''} [P21]
\end{quote}

This seemed to satisfactorily discharge developers' responsibilities---no participants described being worried that they might face consequences for inadequate privacy policies. This confirms evidence of 1,500 Alexa skills found with a privacy policy that was used by another skill~\cite{edu2021skillvet}. Reused privacy policies very often lead to the privacy issues mentioned in Section~\ref{subsec:issues}, as they often lack clear traceability to actual privacy practices. In fact, most of the 1,500 reused privacy policies of skills found in the wild had incomplete or broken traceability with their  privacy practices~\cite{edu2021skillvet}.

When asked more broadly about their use of data participants always saw themselves on the right side of a line, often believing that what they had collected was not personal enough to `count': \textit{``But it's not really personal information, right? [..] I never consider it as personal information.''} [P09]. In particular, the terminology and architecture of the Alexa permissions system caused some confusion here; not all Alexa `permissions' reveal personally identifiable data, and any information available to developers has already been collected by Amazon: \textit{``I mean I'm only using the email address and [people using Alexa have] already given their email address to Amazon or Alexa, so I'm only using that.''} [P12]. Unlike smartphone permissions that relate to device \textit{functionality} through device sensors (and that developers have more experience with, particularly as users), Alexa permissions operate instead at the level of \textit{information} (e.g. the user's postal address), potentially causing further misunderstandings. Relatedly, Alexa permissions involve the disclosure of information about the user \textit{from Amazon} to the developer, rather than directly to a developer from the user.

Professional developers universally framed this in terms of compliance: \textit{``As a general of thumb, I don't want your personal data. It just opens me up to too much risk. As little personal data you can give me is better in order to get the job done. Companies want to know everything and we push back quite strongly on that.''} [P23]

Ironically, the poor user experience of many Alexa operations around data access meant that data minimization was often the best strategy to avoid the large drop-off in engagement that would be caused by requiring users to link accounts or grant permissions: \textit{``I think that in reality [we] struggle to get more than one percent of people to link their account''} [P21], with requests for permissions also resulting in similarly reduced engagement.

Where account linking was required --- e.g., where users had existing service accounts or in the case of skills being used as a cost-saving measure for product support as described in Section~\ref{results:monetisation} --- data and interactions collected through Alexa tended to draw on and supplement off-platform profiles. In this sense, skill permissions were less relevant, as these data had already been collected off-platform, with user interactions used to enrich external profiles and analytics. This is likely to increase given the introduction of Alexa advertising IDs after the conclusion of the interviews.\footnote{\url{https://developer.amazon.com/en-GB/docs/alexa/custom-skills/policy-requirements-for-an-alexa-skill.html} (Section 5, accessed 16/03/2023).}

\subsection{Security, Testing, and Certification}
\subsubsection{Security Mechanisms Used}
\label{results:security:mechanisms}
When asked about security, participants also referred to the insensitivity of the data they held as a justification for not having given additional consideration to securing their voice apps. In contrast to privacy, however, when we asked about security, developers often pointed to the fact that their voice app was hosted on a platform-provided service (e.g., AWS). This meant that it was the platform's responsibility to take steps towards ensuring security and applying any pertinent security mechanisms, and that the liability was not with the developer: \textit{``I try to keep everything within the AWS Cloud. It's easier, it's straightforward, and I don't have to worry as much about security''} [P18].

Beyond this, developers had generally not considered whether they should implement additional security measures or what those would be: \textit{``We do use two-factor authentication and stuff to try and make sure that you shouldn't be able to get into it unless you're supposed to [..] I mean, I've been curious if there are other things that we should be doing that would be better''} [P21]. This situation is understandable given that the Amazon-provided security requirements for most Alexa skills only mention verifying that requests come from the right Alexa skill, which only requires checking that the provided skill ID is correct.\footnote{\url{https://developer.amazon.com/en-US/docs/alexa/custom-skills/security-testing-for-an-alexa-skill.html} (Accessed 23/05/2023).}

\subsubsection{Testing} When it came to testing, there was a distinct split in perspectives between hobbyists and professionals. Professionals saw a voice app's release as a milestone in, rather than the end of, the development process: \textit{``I think at the point that you release, you're about halfway there, generally. Maybe you're 30\% of the way there, maybe 70\%, it depends on how good a job it is and how big and varied your audience of users is gonna be''} [P22]. 

As part of this, a large amount of effort and resources were dedicated to catching errors through different logging and testing methods post-release: \textit{``[we] log errors that are not just technical errors, but usability errors. And then it just gives you a list of how many of these errors you show it [are] showing up in [the skill] in the last week and you just kind of go through the top ones''} [P21].

User testing for hobbyists was usually carried out with friends and family: \textit{``My kids are young, my oldest is 13. So I would sit down and discuss with them what I'm trying to do, and usually, they have good creative ideas or they will criticize''} [P18]. For professionals, user testing at an early stage was seen as crucial and their resources allowed them to recruit a more representative pool of testers. More experienced developers were able to use this knowledge to design voice apps that were less susceptible to usability errors or edge cases in the first place: \textit{``We're getting a lot better at building the edge cases into the `happy path' [..] we know that user will probably have the inclination to fall off here, so we might build some guard rails in to prevent that from happening''} [P23].

\subsubsection{Certification is Inconsistent} 

Testing spilled over into the certification process, with hobbyist developers often seeing certification as a means to \textit{``get feedback from the Alexa developer certification team''} [P01] on the quality of their code. In general, these participants did not have to go through the process often enough to identify systemic issues, often attributing problems to their own work. There was a general consensus amongst all developers that certifying ordinary `run of the mill' functionality proceeded largely without issue but that complicated or unusual functionality was much more likely to generate problems.

Despite this, the amount of time taken for Alexa certification (re)submissions still varied considerably: \textit{``sometimes it goes most smoothly, it only takes one or two days, and sometimes it takes them two weeks''} [P17]. Hobbyists were generally more tolerant of this as they lacked the time pressures associated with launching new business features. This was part of a wider theme where the entire certification process was seen as arbitrary and inconsistent, with certification teams rejecting updated skills because of functionality that had previously passed certification: \textit{``you say `well it's been like that for two years, why have you only just noticed this now?', and then you end up having an argument with them and they're like `oh well it's always been against the rules' and you're like `then why have you allowed it through certification like 20 times?!''} [P21].

Another time that this came into the spotlight was when skills were published simultaneously in different regions. Each region's team would have to independently certify the skill, and they may not agree on whether the skill was ready for publication: \textit{``France would reject the certification based on a certain point, no other territory would have picked up on that, but then Mexico would have an issue with something else that no one else had a problem with. There were some issues that were consistent [..] but I found inconsistencies between different markets''} [P22]. Each time a skill is changed as part of this process it might have to be resubmitted in each market, and each submission might then take several weeks to be processed by Amazon.

More dynamic voice apps that utilize a variety of unpredictable data sources add an additional layer of problems, as they might be tested at any time. When certifying P15's booking app \textit{``[Amazon] wouldn't commit to when they go and test the application [..] they [might not] get the expected flows because of the state of the backend data, because I had other users using a variety of different interaction mechanisms to go and actually book those resources''} [P15]. This meant that a member of the certification team might test the voice app at a time when nothing was available to book and refuse certification as it did not fulfill its stated function.

Finally, professional developers believed that more popular Alexa skills were subject to more stringent certification procedures: \textit{``we go through enhanced certification processes [..] if I log into a non [company] developer console, write a skill, and stick it out there I think it honestly gets a couple of automated checks and as long as it passes that I wouldn't be surprised if it just goes live [..] those skills just don't get any usage, they don't get any traffic pointed to them. So the process for [skills with] more traffic is also more intensive''} [P21]. Developers with high-profile skills were also able to skip steps like auto-enablement, allowing users to invoke their skill without explicitly enabling it in the companion smartphone app.

\subsubsection{Attempts to Finesse Certification}\label{results:finessecertification} As a result of the problems described above, we observed several strategies used by developers to mitigate friction in the certification process. Most professional Alexa developers were able to leverage internal contacts at Amazon to expedite certification or access extra functionality: \textit{``we do have people within Amazon that can be very useful and kind to us [..] if you've got a couple of people's email addresses in Amazon, you can sometimes wiggle it. We try not to do it too often.''} [P23]. These contacts were often developed at in-person conferences and events that hobbyists lacked the awareness, time, or resources to attend:

\vspace{-8pt}

\begin{quote}
\textit{```At one of the developer events I was having a conversation with someone and he said, `I'm from Amazon', and he said `what skills are you working on?' and I mentioned [celebrity] and it was really funny, it was like, alright, okay, come out of the muggle space let's have a grown-up conversation. And at that point I was given the email address and phone number of a contact [at Amazon]''} [P22].
\end{quote}

Across all developers, it made sense to try and minimize the number of certification passes required by a voice app. This was achieved by crafting updates that did not change the front-end interaction model, thus circumventing recertification: \textit{``I try to create just a single intent and have all the possible variations as slots in that specific intent only, so that that allows me to kind of short circuit Alexa's NLP model''} [P04], \textit{``if you can get away with doing a back end change you always would, or [..] putting in dynamic entities or something so that you can do it at runtime rather than having to update the interaction model''} [P21]. 

While the ineffectiveness of voice app certification is a known issue~\cite{cheng2020dangerous, lentzsch2021hey} (see Section~\ref{subsec:issues}), this shows \textit{how} and \textit{why} developers avoid following the spirit of the process. It also highlights a major difference with development for smartphone platforms: instead of app code being hosted by the platform and distributed through the skill store,\footnote{While many smartphone apps do have functionality based in the cloud, all such apps \textit{must} have code hosted on-platform.} forcing recertification when this changes, popular voice apps do not have any code running on the user's physical device.

\subsection{Challenges in Designing for Voice}
\subsubsection{Asymmetric Access to Knowledge}
Smaller developers without connections often struggled when seeking guidance to develop voice apps: \textit{``I didn't have anyone around me that had even begun to write an Alexa skill, I could not ask anyone [..] I use a lot of online resources like YouTube videos and forums and send mails to people and try to fill that gap in the local community''} [P09]. This divide also covered knowledge sharing, with in-person events as a key way to exchange best practices alongside the creation of a dedicated team within an organization: \textit{``we've got a community of practice across the whole company [..] where people share updates from their projects or lessons learned''} [P24].

\subsubsection{The Problem of Skill Names and Discoverability}
A recurring way in which the asymmetric access to knowledge showed in the interviews was around the poor discoverability of voice apps compared to those on smartphones and the web: \textit{``there's no icon on your desktop, there's no icon on your phone to press, so it instantly falls out of your brain''} [P23].

The first hurdle many faced was what to name their app. The obvious choice, particularly when the voice app was an extension of an existing service, was often not the best: \textit{``I figured I would go with [website name] because branding, right? So you know, keep the same as my website, that kind of stuff. Horrible mistake.''} [P13]. As these names had not been initially chosen with a voice in mind, there were often problems with recognition, they felt unnatural to ask for, and could also conflict with existing voice apps --- invocation names do not have to be unique. This can lead to security issues as mentioned in Section~\ref{subsec:issues} such as skill impersonation, by creating specific malicious skill invocation names, reusing the same as or (phonetically) similar  names to other skills~\cite{kumar2018skill,zhang2019dangerous}. In fact, over a quarter of Alexa skills in the wild have names identical or phonetically similar one another~\cite{edu2021skillvet}.

Taking this a step further, professionals spoke about trying to embed as much meaning into voice app names as possible, so that people knew what the app was about before they had even interacted with it: \textit{``they really don't work if you have to explain before you start the conversation how it works, you need to be able to introduce people to the things that they can do as part of the conversation. And so when I was thinking about the first [voice app] to build, I said, well, something where you don't even have to explain anything''} [P21]. In general, anything more than a few sentences of introduction was seen as too long.

Because of this poor discoverability, rather than generating buzz themselves, voice apps needed to be promoted in order to become popular (\textit{``if you build it, they won't come''} [P23]). Successful apps were seen to be those that solved a need in a very focused way: \textit{``So it's really common with digital product design to just throw more features in, make it better by putting more in, more is better as a general thing [..] voice is the other way around, it's completely the opposite, and you have to be super focused''} [P22]. As described above, P23 considered voice SEO as an alternative means of making existing content accessible rather than a dedicated voice app.

\subsubsection{An Evolving Development Landscape}\label{results:evolving}
The interviews revealed a landscape of best practices, tooling, and paradigms that were in constant flux. As mentioned in Section~\ref{results:moneyandpower}, the decision by Google to move from conversational actions to app actions represents a shift in the conceptualization of voice apps that had a big impact on some of our participants. While the shift to app-actions removes the duplicated effort to maintain separate smartphone and voice apps that fulfill the same purpose, for  developers using middleware to create cross-platform voice apps it \textit{increases} the workload by requiring the \textit{creation} of an Android app.

Talk about these middleware platforms themselves revealed their significance and reinforced the impression of a still-evolving landscape: \textit{``There were three main [middleware] platforms a year ago, one of them got bought by Walmart [..] which I guess is a signal of how important conversation design has become to Wallmart as a business---they just decided to buy one of the three main players''} [P22]

And while the technology driving the development of voice assistants had been benefiting from largely incremental gains in recent years, there were still breakthroughs, as we discuss in more detail in the next section, that demonstrated how the field has not yet stabilized:

\vspace{-8pt}

\begin{quote}
\textit{``Absolutely guaranteed for the next 12 months it's all going to be generative AI, it's all going to be GPT-3 and how we can use GPT-3 [..] in half an hour today for each intent I've created about 20 utterances to train that intent just by asking GPT-3 to write me a training set for a chatbot for an intent. That took me less than an hour. There's only 20 utterances but ... to do it manually, that would have been probably two days''} [P22].
\end{quote}

\section{Discussion}
\subsection{Differences With Other Platforms}
\label{subsec:disc-diffs}
Our results point to a number of ways that voice assistant platforms (and thus development for them) differ significantly from other platforms such as smartphones and the web: 

\vspace{-5pt}

\begin{enumerate}
    \item Options for monetization are fundamentally different and reflect deep-seated problems around discoverability in voice interfaces, increasing incentives for developers to find alternatives that violate platform policies and pressure to beta-test new functionality in exchange for support;
    \vspace{-3pt}

    \item Certification is seen as inconsistent, arbitrary, and is often handled opaquely on an informal/ad-hoc basis, leading to reluctance to patch vulnerabilities in voice apps and encouraging circumvention;
    \vspace{-3pt}

    \item Alexa permissions follow a different paradigm to those on other platforms, leading to developer uncertainty about whether they were using personal data or not;
    \vspace{-3pt}
    
    \item Developers are made liable for many privacy and security matters and given little support by vendors, leading to GDPR violations and excess data use;
    \vspace{-3pt}

    \item Voice apps need to be designed very differently to graphical experiences, but security, privacy, and other best practices are often shared within closed communities rather than publicly;
    \vspace{-3pt}

    \item The development landscape changes fast enough to prevent the widespread formation and dissemination of security, privacy, and other best practices.
\end{enumerate}

We discuss the implications of these differences further in the following subsections.

\subsection{Compliance and the Infrastructure of Voice App Publishing}\label{sec:compliance-infra}
On several topics around data use and privacy policies it was clear that developers felt they had been left to their own devices when making privacy and security decisions, with little support from platforms and sometimes even confusion over where privacy and security requirements originated from.

To better understand the situation we can consider the challenges developers face through the role that platforms play as infrastructure. Drawing on~\cite{10.1177/00027649921955326}, we see that voice app platforms possess many key characteristics of technical infrastructure. To users, the privacy, security, and certification mechanisms used are almost completely invisible. They are also fundamentally \textit{relational}: while for developers the certification process was a task primarily of design and \textit{articulation work}~\cite{10.1111/j.1533-8525.1988.tb01249.x} to demonstrate compliance to certification teams, For platforms certification represented a balancing act between reducing spending and minimizing legal liability and the perceived need for additional regulation. We unpack these in more detail below before discussing how the values that underpin the infrastructure affect developer behavior.

\subsubsection{Reducing Legal Liability}
While infrastructure is usually seen as embodying the standards of the sector~\cite{10.1177/00027649921955326}, which in this case would include the GDPR~\cite{GDPR16}, CCPA~\cite{CCPA2018}, and/or the new EU Cyber Resilience Act,\footnote{\url{https://digital-strategy.ec.europa.eu/en/library/cyber-resilience-act} (accessed 21/09/2023)} what we found was a situation that made enforcement action by regulators exceedingly difficult. Firstly, the architecture of voice apps as discrete packages of third-party software means that developers in the US, EU, and jurisdictions with similar regulations are considered data controllers (CCPA: ``businesses''), and as such are legally responsible for properly collecting and protecting data and any violations that arise. Secondly, the EU Cyber Resilience Act ensures that security features are independently verified by establishing a framework for secure development, a situation that can hardly be met when voice apps run in the cloud and with the re-certification issues discussed in Section~\ref{results:finessecertification}.

At the same time, platforms offered little support to developers to navigate the privacy and security responsibilities placed onto them by these architectures, wary of being held liable for violations where developers had followed platform instructions. We saw this in Sections~\ref{results:security:mechanisms} and~\ref{results:privacy:use-policies} where developers felt liable for privacy and security matters, understanding that privacy and security were their responsibility and not that of platforms. Developers noted a lack of guidance from platforms on how they should make decisions around privacy and security and often lacked the expertise/confidence to tackle these issues themselves. This in turn led to developers relying on others who were seen to be more knowledgeable wherever possible, copying privacy policies and assuming that security issues would be handled by their hosting provider, even when this was the platform they were using.

\subsubsection{Reducing the Perceived Need For Regulation}
Within the wider context of data protection enforcement, this approach by platforms of pushing liability onto developers makes sense. Smaller skills are very unlikely to be subject to attention from academics, journalists, or regulators, whereas platforms \textit{are}. This leads to an optimal strategy of accepting as little liability as possible without prompting additional regulation of platforms (e.g., the unsuccessful US Platform Accountability and Transparency Act,\footnote{\url{https://www.congress.gov/bill/117th-congress/senate-bill/5339} (accessed 28/04/2023).} and as has happened with the EU Digital Services Act for social media --- DSA~\cite{DSA}). To achieve this, we saw speculation that Amazon subjected more popular Alexa skills (i.e., those that are more likely to attract attention from academics, the press, and regulators) to more scrutiny, taking a calculated approach to moderate the platform as cheaply as possible. Contrast this with prior work that found hundreds of skills with inadequate or nonexistent privacy policies that remained so even after being reported~\cite{edu2022measuring}.

As to why existing regulation like the DSA does not incentivize small and medium developers to follow good data practices, there are very few sanctions against developers of small and medium-sized skills with bad data practices. If vendors like Amazon are not checking that privacy policies are up to date, and data protection authorities are struggling to enforce complaints against even the biggest companies,\footnote{\url{https://www.wired.co.uk/article/gdpr-2022} (accs 28/04/23).} then for smaller companies the chances of being enforced against are very slim. As well as developers who do not know better or have made a calculated decision, this includes those who might seek to take advantage of the lack of regulator attention (e.g., by circumventing permissions frameworks by asking for personal data in-conversation~\cite{edu2021skillvet}). In effect, by pushing liability downwards in the manner currently seen big platforms have created an environment that allows for poor data protection practices with a low risk of enforcement action for any party, possibly elucidating prevailing substandard methodologies and insufficiency of attention towards privacy concerns well known to platforms~\cite{edu2022measuring}. At the same time, platforms are able to claim that they have systems in place (and dictate the entire certification process from start to end), to counter arguments that tighter regulation is needed.

\subsubsection{Trickle-down Values}
Finally, we examine how the above values that underpin voice app infrastructure further influence developer behaviors by posing a counterfactual. The placement of compliance at the top of platforms' agendas in turn leads to developers prioritizing approaches perceived to be compliant over those that are usable and accessible. Consider the difference if Amazon or Google provided a privacy policy template that had an overview in the style of \textit{Terms of Service; Didn't Read}, which provides a bullet point summary of website terms and conditions.\footnote{\url{https://tosdr.org} (accessed 28/04/2023).} Future work could explore this design space, which would benefit users through more easily understandable privacy policies and developers through greater clarity.

The irony is that from the perspective of regulators, the systems that \textit{are} in place actually fail to properly promote compliance. We describe in Section~\ref{results:finessecertification} how some participants had developed ways to avoid re-certification and highlighted some of the strategies used, the existence of which had been hinted at in prior work~\cite{cheng2020dangerous,lentzsch2021hey}. Combined with prior work on circumvention of the permissions framework and collection of personal data in-conversation~\cite{edu2021skillvet, 10.1145/3543829.3544520,guo2020skillexplorer,young2022skill}, it is now clear that circumvention of both the permissions and certification processes on the Alexa platform is not just theoretically possible but present in the wild. Given the problems we found around monetization and the restrictive advertising policies on platforms, there is an incentive for developers to make use of these loopholes  to make their skills more financially viable.

\subsection{Developers' Lack of Security Awareness}
Despite encouraging participants to discuss their security practices during the interviews by asking the questions described in Section \ref{sec:analysis}, it was clear there were widespread knowledge gaps and difficulties identifying and articulating security practices/behaviors. The appeal of platform-provided services as a catch-all answer to issues of security is likely driven in part by platforms themselves marketing their hosting services as secure (e.g., AWS includes marketing lines such as ``All data is stored in highly secure AWS data centers''\footnote{\url{https://docs.aws.amazon.com/whitepapers/latest/aws-overview/security-and-compliance.html}(accs 18/09/23).}),  though future work should investigate this further.

There is ongoing work in the nascent area of standards, guidelines, and best practices for voice application development. For instance, the Open Voice Network\footnote{\url{https://openvoicenetwork.org/our-work} (accs 11/09/23).} is a non-profit community of the Linux Foundation pursuing technical standards in security and privacy specific to voice apps such as user privacy and voice data security which, if adopted, would represent a significant step towards improving voice developers' security and privacy knowledge/practices. Research is also developing more usable security and privacy mechanisms in voice assistants~\cite{zhan2023privacy}, which could be provided by the platform in the future to help with this endeavor.

\subsection{Financial Incentives For Developers}\label{sec:discussion-financial}
There are often tensions between app stores and those who develop for them, particularly around commission rates and moderation. Third-party software represents a significant selling point of compatible devices, and the health of these platforms depends on the extent to which developers can support themselves financially. Throughout the interviews, we saw how this is incredibly difficult on voice assistant platforms. Voice interfaces offer poor discoverability, limiting user growth in the absence of expensive marketing campaigns. This increases the pressure on developers to accept promotions from Amazon in return for implementing riskier features that would not otherwise be to their advantage.

At the same time, platforms restrict the ability of developers to support themselves through advertising and provide unoptimized in-skill payment flows, which negates the benefit of Amazon paying developers ``up to 80 percent of the marketplace list price''\footnote{\url{https://developer.amazon.com/en-GB/docs/alexa/paid-skills/payments.html} (accessed 28/04/2023).} for purchases made through their skills. This is higher than comparable platforms but ignores the fact that developers are driven to off-platform payment methods. This is less problematic for large services (e.g., Spotify), which already have established web and app payment flows, but is much more difficult for developers whose creations only exist on voice assistant platforms. Developers are also able to earn commission through the Amazon affiliate program---by placing items into the user's Amazon shopping cart---but prior work has shown that users are extremely hesitant to purchase items through voice assistants~\cite{abdi2019more}.

Finally, the cash injections from Amazon which kick-started the creation of third-party skills have slowly dried up, leading to a \textit{de facto} end to the emergence of new skills that are financially viable. From our data, it is difficult to predict whether the Alexa platform can maintain an equilibrium with its current developers, or whether the unique circumstances around voice apps mean that the marketplace requires financial support from platforms in order to continue.

As previously mentioned, since the date of the interviews there has been a change to the Alexa advertising policy that permits developers to track users using an advertising ID similar to that used on mobile platforms. While this may be a welcome additional source of income for skill developers, the fact that skills can not provide general advertising unrelated to the skill means that this will not be as effective on approach as on mobile platforms where advertising alone can cause an app to be economically sustainable. As Alexa developers already receive a persistent user ID this is unlikely to affect services that already collect user data for use off-platform---these IDs are pseudonymous until a user links their off-platform account through the skill (cf. account linking in Section~\ref{results:privacy:use-policies}).

\subsection{Recommendations \& Implications}
\paragraph{For Platforms}
A clear first step would be to review certification processes and staffing with the aim of providing greater consistency and clarity over certification response times and clearer communication when an app fails certification. As part of this, procedures and training should be systematized in order to make decisions less subjective (and thus less variable), and there should be a procedure for changing content policies that notifies affected developers and gives them ample time to modify already certified apps. When it is not possible to do this in advance, re-certification of existing code should take place in parallel with new changes, come with a grace period, and not block app updates. Finally, newly introduced functionality should be cleared with certification teams \textit{before} being made available to developers.

\paragraph{For Regulators}
The way that the certification processes have been designed and subsequently evolved means that self-regulation is not appropriate to ensure user privacy and security (cf. Section~\ref{sec:compliance-infra}). Tighter regulation is needed to ensure that platforms are not able to dodge responsibility for privacy and security.

Where platforms dictate the processes by which developers obtain and/or process data, such as by forcing them to use a given permissions/consent framework, then they should be obligated to provide guidance on using those processes. Platforms should also be obligated to regularly re-verify the compliance of voice apps on their platforms (e.g., by checking that they have privacy policies), notifying non-compliant apps and giving them a fixed amount of time to address any issues. This is similar to social networking sites, where users are responsible for the speech that they post, but platforms are also required to moderate that content.

Several of these issues stem from architectural decisions and could be most cleanly addressed by platforms requiring that code be hosted under their control (e.g., AWS or Firebase), as happens with smartphone apps. This would allow for greater automated analysis of apps through static and dynamic analysis of source code and mitigate functionality changes without triggering re-certification.

Regulators could also make use of the growing open-source tools being developed by the academic community such as SkillExplorer~\cite{guo2020skillexplorer}, SkillVet~\cite{edu2021skillvet}, and SkillDetective~\cite{young2022skill} to conduct independent audits on voice apps in the wild in voice assistant marketplaces/ecosystems. This could provide some evidence of the state of affairs without necessarily involving voice assistant platforms.

\paragraph{For Researchers }
Our findings lead us to reflect on the methods commonly used to conduct research on the privacy and security of voice apps. The potential existence of enhanced certification processes for higher traffic skills, as well as inconsistencies with certification within platforms, mean that popular methods  within the research community may not fully capture the situation on the ground.

The main problem is one of scale; it is impossible to know how many users or activations a skill has unless you are the developer of that skill. The results of the interviews suggest that the currently observed privacy problems with skills may be concentrated on those with fewer activations, with more ad-hoc and opaque processes used to hold more popular skills to higher standards. The scraping and interaction~\cite{guo2020skillexplorer,edu2021skillvet,young2022skill} or sock puppet~\cite{cheng2020dangerous} approaches used in prior work treat skills as equal regardless of user engagement, meaning that the real impact may be overstated. 

The interview data also suggested that in terms of accessing personal information, data account linking was potentially more useful than the permissions frameworks used by voice assistant platforms. However, work in this space has previously only focused on permissions, as data usage via account linking is exceptionally difficult to track at scale --- privacy policies are the only publicly available information, and these are often vague or inaccurate~\cite{guo2020skillexplorer,edu2021skillvet,young2022skill}.

\subsection{The Potential for LLMs and Generative AI}
Since exploding in popularity in 2022 with the release of ChatGPT, large language models (LLMs) will undoubtedly influence the evolution of voice assistants over the next decade. The research community has yet to fully understand the risks associated with LLM use; meanwhile the Italian Data Protection Authority (Garante) temporarily banned ChatGPT over its use of personal data, and the European Data Protection Board has set up a ChatGPT task force \cite{Supantha_2023}. 

We observed in our interview how LLMs had already begun to change the development process, with P22 using ChatGPT to generate sample utterances for intents (Section~\ref{results:evolving}). The conversational nature of these products positions them perfectly to disrupt the voice assistant market. Research has shown how voice apps already circumvent platform policies on data collection by asking for data verbally in-conversation rather than using permissions frameworks~\cite{edu2021skillvet,guo2020skillexplorer,young2022skill}, and the ability to easily generate natural dialogue asking for personal data could see this practice increase, mirroring  work on self-disclosure and computer-originated dialogue~\cite{moon2000intimate}.

\subsection{Limitations}
There were only a few participants who had successfully managed to gain hands-on experience around monetization and beta functionality. As such, the sample size for these aspects of the results was smaller, although, as we note in Section~\ref{sec:methods}, these participants tended to be the most experienced of the sample, which meant that their interviews usually led to richer, more nuanced, and more informed answers, opinions and explanations. This was very useful for the purposes of a qualitative study like ours, which precisely aimed to explore and understand in-depth participants' perspectives in order to generate hypotheses that can then be tested in follow-up quantitative studies; as with all qualitative work, our results represent the experiences of our sample.

\section{Related Work}
\label{sec:related}
\subsection{Voice App Developers: Existing Gaps}\label{relatedwork:gaps}
While a small amount of research has examined the development of voice apps using technical and policy framings, none have looked at the voice app ecosystem from the perspective of \textit{developers}. By considering experiences from actual developers, \textbf{we address a gap in the literature that prevented a deep understanding of the circumstances that often lead to the creation of problematic skills}; to the best of our knowledge, this work is the first to offer a root-cause analysis of the underlying security and privacy issues of the development process, unveiling new issues that have been limiting the widespread proliferation of voice assistant technology.

This existing work has explored how voice app developers are given legal and other responsibilities without the corresponding control over how they can access user data~\cite{seymour2023legal}, how Alexa certification works behind the scenes with human and automated testers~\cite{10.1145/3478101}, and how the Alexa and Google certification processes miss many policy-violating skills and actions~\cite{cheng2020dangerous}. These works explore how theoretical adversaries could potentially game voice app certification and permissions, offering insights into how market operators should fix vulnerabilities in the certification and consent mechanisms.

\subsection{Voice App Vulnerabilities}
Thousands of voice apps have been developed by third parties \cite{edu2022measuring}, but previous research has not explored the creation processes or practices of developers. Instead, studies have focused only on analyzing the vulnerabilities and privacy policies of these skills. For example, a study of 77,957 voice apps found that 5\% of these applications could perform sensitive actions using hidden commands~\cite{shezan2020read}. Others~\cite{guo2020skillexplorer, zhang2019dangerous} have shown that some skills may continue to listen even after the user has commanded them to stop listening. Cheng et al.~\cite{cheng2020dangerous} demonstrated that developers could exploit the discovery mechanism of voice platforms to increase the popularity of their apps. 
In \cite{kumar2018skill}, the authors show how developers could use the inherent errors in the NLP algorithms and words that are often misconstrued to create malicious skills and exploit the ambiguity in the skill invocation name method. The authors ``hijacked'' the skill invocation names of more than 50\% of five randomly sampled vulnerable target skills.
More recently, Young et al. \cite{young2022skill} tested  54,055 Amazon Alexa skills and 5,583 Google Assistant actions, and  found that 6,079 skills and 175 actions violated at least one policy requirement. This is not surprising considering the findings of our study, and highlights the need we emphasize in the discussion to better support developers in this space. 

\subsection{Certification and Violation of Policies}
While exploring certification mechanisms, many studies have exposed gaps in the process and argue that many of the published voice apps violate their own privacy policies and platform requirements~\cite{10.1145/3543829.3544513}. Some studies~\cite{cheng2020dangerous,lentzsch2021hey} argue that the certification process is inconsistent, and verification relies heavily on the information provided by the developers instead of actually testing voice apps. Many studies have reported that developers often violate policies and requirements, causing concern. For example, Cheng et al.~\cite{cheng2020dangerous} submitted 615 policy-violating voice apps to the Alexa and Google Assistant stores. All Alexa skills and 39\% from Google Actions passed certification and were made publicly available. They suggest that this may be due to too much responsibility placed on developers or unrealistic expectations.

Other studies~\cite{guo2020skillexplorer,lentzsch2021hey} have shown that some skills collect sensitive data outside of the permission ecosystem or data that do not align with the voice app description. Moreover, Edu et al.~\cite{edu2021skillvet} discovered that some skills could bypass permission and breach their privacy policy despite the privacy settings being available. Meanwhile, Le et al.~\cite{le2022skillbot} found that while children's skills undergo an intensive verification process, some skills still expose children to inappropriate content and amplify or spread misinformation~\cite{edu2023misinformation}. There is also some evidence~\cite{cheng2020dangerous,lentzsch2021hey} that developers could make code changes after certification and collect more personal information. 

Regarding privacy policies, several studies~\cite{alhadlaq2016,liao2020measuring,guo2020skillexplorer,lentzsch2021hey,edu2022measuring} have highlighted issues around missing or broken links. For example, Liao et al. found 17,952 skills and 9,955 actions with broken links on Amazon and Google platforms, respectively~\cite{liao2020measuring}. In other instances, they found that some developers duplicated privacy links between several skills. Previous studies~\cite{guo2020skillexplorer,liao2020measuring,lentzsch2021hey} have reported instances where developers requested private information without reporting it in their privacy policy. Instead, they directly asked users for private information during conversations. After reviewing various skills, Lentzsch et al.~\cite{lentzsch2021hey} revealed that around 23.3\% of privacy policies did not address the data collected through the skills. In 2017, 70\% of skills with privacy policies did not include any information specific to voice apps or systems~\cite{alhadlaq2016}. Despite these findings, no previous studies have investigated the practices of developers from the \emph{developer's perspective}, e.g., in creating privacy policies for voice apps.

\subsection{Developers' Security \& Privacy Practices in Other Platforms}
As described in Section~\ref{relatedwork:gaps}, while a plethora of security and privacy literature exists concerning developers in similar platforms such as the mobile app ecosystem, no prior studies have focused explicitly on the experiences of voice app developers. Within the domain of mobile app development, extant research reveals that developers are prone to engaging in risky practices, including the disregard of security implications~\cite{acar2016you}, code reuse from online sources~\cite{kim2004ethnographic}, excessive permission requests~\cite{felt2011permission,felt2011android,Tahaei2023Stuck}, privacy policy violations~\cite{slavin2016toward,yu2016can,andow2020actions}, neglect of Transport Layer Security (TLS)~\cite{fahl2012eve,fahl2013rethinking}, and improper usage of cryptographic APIs~\cite{egele2013empirical,mutchler2015large}. 

Developers face several challenges when it comes to regulatory compliance, having trouble understanding privacy requirements~\cite{li2018coconut,bednar2019engineering,senarath2019will}. Even when they know these requirements, they may not know how to use the provided SDK to implement the necessary privacy settings~\cite{green2016developers,li2018coconut,li2021developers,tahaei2022charting}. Some developers expect platform  support in this area~\cite{li2021developers}, while others believe it is a shared responsibility between them and platforms~\cite{li2021developers,tahaei2022charting}.

Bednar et al.~\cite{bednar2019engineering} argue that though developers may intend to build secure software, implementing security requirements frequently presents daunting technical challenges. This is compounded by inadequate usability of tools and libraries, introducing system vulnerabilities~\cite{green2016developers}, lack of secure alternatives~\cite{li2021developers}, dearth of guidance regarding optimal configurations~\cite{tahaei2022charting}, and inscrutable documentation~\cite{abdalkareem2017developers,meng2018application,patnaik2019usability}. To assist developers, prior works have proposed diverse interventions, such as warnings~\cite{gorski2018developers}, plugins~\cite{nguyen2017stitch}, and the simplification of  APIs~\cite{indela2016helping,kruger2017cognicrypt}. However, studies have not looked at the challenges that voice app developers face and how they impact security and privacy. Voice apps have key differences with other platforms (cf. Section~\ref{subsec:disc-diffs}) around monetization, liability, certification, and personal data collection that bring different challenges for developers. 

\subsection{Usable Voice App Privacy \& Security}
There is extant literature on user perceptions of the privacy and security of voice assistants, with data collection and scope, `creepy' device behavior, and violations of personal privacy being reported as major concerns~\cite{10.1145/3170427.3188448,10.1145/3313831.3376529,10.1145/3449119, 10.1145/3359161,cho2019hey,moon2000intimate,lau2018alexa}. Research has shown that users often have incomplete mental models of how voice assistants work, leading to different (and often incorrect) perceptions about what data is shared and with whom. Common mistakes include assuming that all voice apps are first-party (e.g. part of Alexa itself), instead of having been developed and ran by third parties~\cite{seymour2023ignorance,abdi2019more,major2021alexa}, and being unable to identify which voice app will be opened for a given command, with this uncertainty leading to a reticence around e.g. making purchases through voice assistants~\cite{abdi2019more, 10.1145/3491102.3517510}. This clearly links to our findings in this paper around the challenges developers encounter themselves around voice app discoverability and the name they choose for them ---that is, the same model that makes it hard for developers to make their voice apps known could be facilitating the mental models users have that everything is the voice assistant, and the challenges developers find to monetize, particularly when it comes to voice app in-purchases, as shown in~\cite{seymour2023ignorance,abdi2019more,major2021alexa}, could be due to this reticence that has been observed from the part of users due to security and privacy concerns. Future work should look more closely at these potential relationships between users and developers and how that may be mediated by the voice assistant platform. 

\section{Conclusion}
In this paper, we conducted a user study that sheds light on the experiences of voice app developers. 
We made significant contributions to the understanding of the challenges they encounter. Firstly, we established a clear connection between issues such as liability and certification with negative privacy and security outcomes observed on voice assistant platforms. Our findings revealed the risks arising from misperceptions regarding the shifting of responsibility between developers and dominant platforms such as Amazon Alexa and Google Assistant. 
Furthermore, we identified critical issues related to monetization, privacy, design, and testing. In particular, we encountered connections between issues raised by developers around difficulties with monetization that include concerns around the {security} of being able to spend money via voice. We confirmed that many of these issues stem from fundamental problems within the voice app certification process. This highlights the need for improvements and streamlining in the certification procedures to address these challenges effectively. Finally, we discussed the implications of our study for various stakeholders, including developers who face risks associated with liability, regulators who need to consider the certification process's shortcomings, and researchers who can benefit from our insights for future studies on voice apps.

\subsection*{Acknowledgments}
This research was primarily funded by EPSRC under grant \emph{SAIS: Secure AI assistantS} (EP/T026723/1). The research was also funded by the INCIBE's strategic SPRINT (Seguridad y Privacidad en Sistemas con Inteligencia Artificial) project with funds from the EU-NextGenerationEU, MCIN/AEI /10.13039/501100011033 and the EU-NextGenerationEU under grant TED2021-132900A-I00.  G. Suarez-Tangil has been appointed as 2019 Ramon y Cajal fellow (RYC-2020-029401-I) funded by MCIN/AEI/10.13039/501100011033 and ESF Investing in your future.  K. Ramokapane's time was supported by EPSRC under EP/V011189/1 and EP/W025361/1.

\bibliographystyle{plain}
\bibliography{main}

\end{document}